\documentclass[aps,prl,twocolumn,groupedaddress]{revtex4-1}
\usepackage{epsfig,amsmath,amssymb,graphics,color,calc}
\usepackage{mathtools}

\begin{document}


\title{Delocalized Glassy Dynamics and Many Body Localization}


\author{G. Biroli\textsuperscript{1,2}, and M. Tarzia\textsuperscript{3}}

\affiliation{\textsuperscript{1}\mbox{Institut de physique th\'eorique, Universit\'e Paris Saclay, CEA, CNRS, F-91191 Gif-sur-Yvette, France}\\ \textsuperscript{2}Laboratoire de
  Physique Statistique, Ecole Normale Sup\'erieure, PSL Research
  University, 24 rue Lhomond, 75005 Paris, France.
  \\ \textsuperscript{3} \mbox{LPTMC,~CNRS-UMR 7600,~Sorbonne Universit\'e,~4 Pl. Jussieu,~75252 Paris c\'edex 05, France}}



\begin{abstract}
We analyze the unusual slow dynamics that emerges in the bad metal delocalized phase preceding the Many-Body Localization transition by using single-particle Anderson Localization on the Bethe lattice as a toy model of many-body dynamics in Fock space. 
We probe the dynamical evolution by measuring observables such as the imbalance and equilibrium correlation functions, which display slow dynamics and power-laws strikingly similar to the ones observed in recent simulations and experiments. We relate this unusual behavior to the non-ergodic spectral statistics found on Bethe lattices. 
We discuss 
different scenarii, such as a true intermediate phase which persists in the thermodynamic limit versus a glassy regime established on finite but very large time and 
length-scales only, and their implications for real space dynamical properties. In the latter, slow dynamics and power-laws extend on a very large time-window but are eventually cut-off on a time-scale that diverges at the MBL transition.    
\end{abstract}


\maketitle
Understanding the inter-play of quenched disorder, interactions and quantum fluctuations has been
a central theme of hard condensed matter for many years. Activity on this topic boomed 
recently, in particular after that
Basko, Aleiner and Altshuler (BAA) showed by using the self-consistent Born approximation that interacting and isolated quantum systems can fail to thermalize due to Anderson localization in Fock space~\cite{BAA}. This phenomenon, called Many Body Localization (MBL), represents a new kind of 
ergodicity breaking transition, which is purely dynamical---indeed it can take place even at infinite temperature by increasing the amount of disorder---and which results from the interplay of disorder, interactions and quantum fluctuations~\cite{reviewMBL,reviewMBL2}.
One of the most surprising results is that even the delocalized phase is unusual in a wide range of parameters {\it already before the MBL
transition}. In fact both in numerical simulations~\cite{dave1,demler,alet,torres,BarLev} and in experiments~\cite{experiments1,experiments2,experiments3} it was found that transport appears to be sub-diffusive and that out-of-equilibrium relaxation toward thermal equilibrium is slow and power-law-like with exponents that gradually approach zero at the transition.
Several works explained this behavior in terms of Griffiths regions, i.e., rare inclusions of the localised phase which impede transport and relaxation~\cite{reviewdeloc1,griffiths,griffithsb,griffiths2}. However, also quasi-periodic $1d$ and disordered $2d$ systems, in which Griffiths effects should be absent or milder~\cite{reviewdeloc1,griffiths2}, do display analogous unusual transport and relaxation \cite{daverecent,dave2d,experiments2,experiments3}. It is therefore important to look for other explanations that might
hold beyond the particular case of $1d$ disordered systems. Moreover, it is interesting 
to complement the real space Griffiths perspective to one directly based on quantum dynamics in Fock-space.  
These are the aims of our work. \\
As a matter of fact, already in ~\cite{BAA} it was argued in favor of a bad metal phase characterized by unusual transport. Even before that, by mapping MBL to Anderson localization on a Bethe lattice~\cite{dot}, it was suggested that the delocalized phase could be non-ergodic, i.e., not fully thermal: in an entire regime of parameters, before the MBL transition, wave-functions could be delocalized but not uniformly spread and could show multifractal behavior. The existence of such delocalised non-ergodic phase 
in the (non-interacting) Anderson model on tree-like structures~\cite{noi,scardicchio,altshuler,mirlin,lemarie} and on related 
random matrix models with long range hopping~\cite{levy,kravtsov,facoetti} 
has been the focus of an intense research activity in the last five years. Although it is still debated 
whether this phase indeed exists for infinitely large systems, it is indisputable that finite-size samples do not
display fully ergodic behavior even far from the localization transition and for very large sizes. In this work, by focusing on the non-interacting
Anderson model on the Bethe lattice as a toy model of MBL~\cite{dot,BetheProxy1,BetheProxy2,scardicchioMB},
we show that these non-ergodic features of the spectral statistics leads to unusual slow and ``glassy'' 
dynamics in a broad region of parameters within the delocalized phase. In particular, in this regime
Bethe lattice proxies for observables such as equilibrium correlation functions and the imbalance display a power-law dynamical behavior completely analogous to the one found in realistic many-body interacting systems. 

As anticipated, we consider Anderson localization on the Bethe lattice, originally introduced and studied in~\cite{ATA}, as a simple framework for 
MBL~\cite{dot,BetheProxy1,BetheProxy2,scardicchioMB}.  This system corresponds to a tight-binding
Hamiltonian for spinless non-interacting fermions, where the quenched disorder is due to on-site random 
energies which are taken as i.i.d. random variables with a uniform distribution between $[-W/2,W/2]$ (we set the hopping $t=1$).
The underlying lattice structure is a random-regular graph~\cite{RRG}. 
In the analogy with MBL, sites should be interpreted as many-body configurations, and on-site energies as extensive energies of a
$N$-body interacting system~\cite{dot,BetheProxy1,BetheProxy2,scardicchioMB}. The two main---drastic---simplifications that we make 
are the following ones:\\ 
{\bf 1.} The configuration space of a many-body disordered quantum system is a very high-dimensional space. For example for Ising
spins it corresponds to a hypercube 
in $N$ dimensions, where $N$ is the total number of spins of the system; MBL can hence be viewed 
as single-particle Anderson localisation on a very high dimensional lattice with correlated random energies. 
By considering Anderson localization on a Bethe lattice as a toy model, we retain the infinite dimensional character 
of the configuration space~\cite{Bethe_dinf} 
but we neglect the correlations between energies as well as the specific structure of the 
hypercube. Note, moreover, that we shall consider a finite connectivity Bethe lattice, whereas the Fock space has a connectivity that increases logarithmically with $N$.\\
{\bf 2.} We are interested in studying averages and correlation functions of local operators in real space. 
In order to figure out how to model such local operators on the Bethe lattice, 
let us focus on the following exemple. 
Here and henceforth we shall consider a random disordered quantum spin-chain, such as the one studied in \cite{BarLev}, as a reference model to explain our procedure. As local observable we take the $z$-component of the spin $\sigma_i^z$. The representation of 
 $\sigma_i^z$ in Fock space is simply $\sum_{\cal C} |{\cal C}\rangle \langle {\cal C}| f_{\sigma_i^z}({\cal C})$ where 
 ${\cal C}=\{\sigma_1^z,\dots,\sigma_N^z\}$ and $f_{\sigma_i^z}({\cal C})$ is equal to the value of $\sigma_i^z$ in the configuration 
${\cal C}$.
The main properties of the function $f_{\sigma_i^z}({\cal C})$ is that it changes in a rapid and scattered way along the hypercube and it is equal to $1$ (respectively, $-1$) for half of the configurations. 
On the Bethe lattice we approximate such complex behavior by a random one by defining local operators as
$
\hat O_{\rm local}=\sum_{\cal C} |{\cal C}\rangle \langle {\cal C}| f_{O}({\cal C})
$, 
where $\cal C$ denotes a site of the lattice (i.e., a proxy for a many-body configuration), and $f_{O}({\cal C})$ is a random binary variable equal to $\pm 1$ with probability $1/2$~\cite{diagonal}.\\
Without loss of generality, we consider the transition induced by increasing $W$ at infinite temperature. It takes place when the states in the middle of the spectrum, i.e., at $E=0$, become localized.
For a Bethe lattice with connectivity three, which is the model we focus on henceforth, this happens at $W_c \approx 18.1$. 
The observables we shall study are the imbalance and the two-point equilibrium dynamical correlation function. 
The former measures whether, say, an initial random magnetization profile converges to its flat thermodynamic average or remains instead inhomogeneous  even at very long times; this 
corresponds to check whether $\sum_i \langle \sigma_i^z(t)\rangle_{\rm rand} ^2/N $ tends to zero or to a positive residual value at long times, where 
\[
\langle \sigma_i^z(t) \rangle_{\rm rand} = \langle \psi_0|e^{iHt}\sigma_i^z e^{-iHt}|\psi_0\rangle \, .
\]
and $|\psi_0\rangle$  is a random initial state (we rescaled time by $1/\hbar$). 
The Bethe lattice counterpart of $|\psi_0\rangle$ 
is a random site $|x_0 \rangle$ 
whose energy is close to zero; following the approximation discussed before, the counterpart of 
$\langle \sigma_i^z(t) \rangle_{\rm rand}$ reads:
\[
\langle \sigma_i^z(t) \rangle_{\rm rand}\equiv \sum_{x=1}^{M}f_{\sigma_i^z}(x)\left|\sum_\alpha'\langle x_0|\alpha\rangle 
\langle \alpha|x\rangle e^{iE_\alpha t} 
\right|^2 \, .
\]
where we have denoted $M$ the number of sites of the Bethe lattice, and $E_\alpha$ and $| \alpha \rangle$
the eigenvalues and the eigenvectors of the single-particle Anderson Hamiltonian.
The prime means that we restrict the sum over eigenstates around zero energy. In a many-body system  
this restriction is automatically enforced by the scaling of the energies in the thermodynamic limit: the states 
that matter physically, even the virtual ones, have all the same intensive energy.  In the model we focus on, which lacks 
this concentration property,  
we have to impose it as a constraint~\cite{microcanonical_Tinf}. \\
Averaging over the disorder~\cite{average}, we thus obtain the Bethe lattice proxy for the imbalance, which reads:
\[
\begin{aligned}
I(t) & = \overline{\Big[ \frac 1 N\sum_i \langle \sigma_i^z(t)\rangle_{\rm rand} ^2 \Big]} \equiv \overline{\sum_{x=1}^{M}\left|\sum_\alpha'\langle x_0|\alpha\rangle
\langle \alpha|x\rangle e^{iE_\alpha t}
\right|^4} \, .
\end{aligned}
\] 
\begin{figure}
\includegraphics[width=0.42\textwidth]{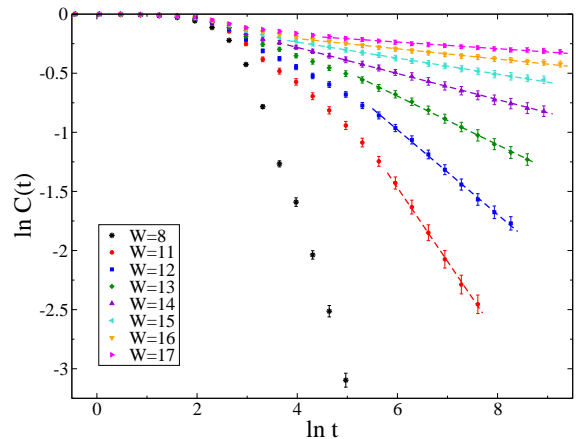}%
\caption{\label{fig:Ct}
Equilibrium correlation function, $C(t)$, as a function of time for different disorder strengths (log-log plot). 
Dashed straight lines highlight the apparent power-law behavior $C(t)\sim t^{-\alpha}$.} 
\end{figure}
\begin{figure}
\includegraphics[width=0.42\textwidth]{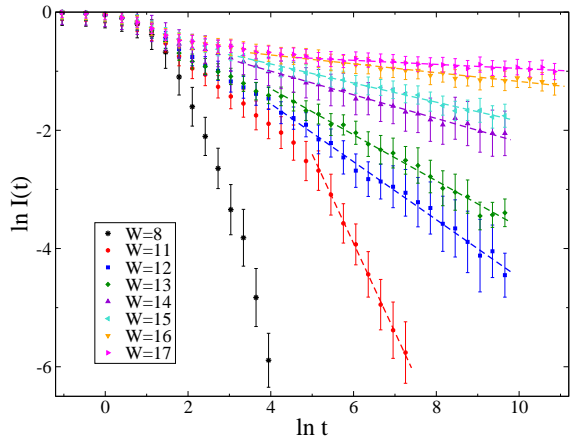}%
\caption{\label{fig:IB}
Imbalance, $I(t)$, as a function of time for different disorder strengths (log-log plot). Dashed straight lines highlight the apparent power-law behavior $I(t)\sim t^{-\beta}$.} 
\end{figure}

\begin{figure}
\includegraphics[width=0.42\textwidth]{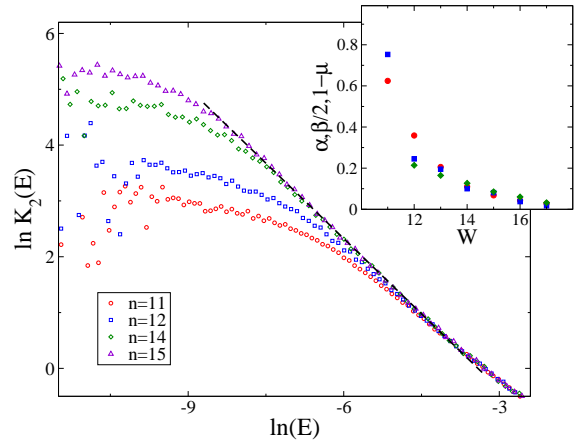}%
\caption{\label{fig:pd}
Log-log plot of $K_2(E)$ for different system sizes $M=2^n$ (n=11,12,14,15). The straight dashed line highlights the apparent power-law behavior $K_2(E)\sim E^{-\mu}$. Inset: Power-law exponents $\alpha,\beta,\mu$ observed numerically as a function of $W$. For smaller values of $W$ our precision on the exponents decreases since we can track the power-laws for less decades. The difference found for $W=11,12$ is likely due to this lack of precision. } 
\end{figure}

\noindent
Note that because of the constraint on the sum over eigenstates the right-hand side of the expression above
is not equal to one for $t=0$. In order to cure this pathology of the model we normalize the previous expression 
by its value at $t=0$.  
 \\
Following the same kind of reasoning, one can define the Bethe lattice proxy for the equilibrium dynamical 
correlation function 
\[
\begin{aligned}
C(t)&=\overline{\frac{1}{2N} \sum_i \langle \left(\sigma_i^z(t)\sigma_i^z(0)+\sigma_i^z(0)\sigma_i^z(t)\right)\rangle_{T=\infty}}\\ 
&\equiv \frac 1 Z\sum_{\alpha,\beta}'\sum_{x=1}^M|\langle \alpha|x\rangle|^2
|\langle \beta|x\rangle|^2 \cos\left[(E_\beta-E_\alpha)t \right] \, ,
\end{aligned}
\] 
where $Z=\sum_{\alpha}'$ is the partition function at infinite temperature. As for the imbalance, we normalize 
by the value at $t=0$ \cite{returnp}.  
\\
These two observables are plotted in Figs.~\ref{fig:Ct} and \ref{fig:IB} for disorder strengths $W=8,\ldots ,17$, for the largest available 
system size, $M=2^{15}$, and by averaging over 64 samples. 
For this value of $M$ our numerical data do not display finite size effects on the time-scales shown in the figures 
(we do observe finite size effects on much larger time-scales or for smaller system sizes, see SI). It is clear from the figures 
that for $W \gtrsim 10$ a regime of slow dynamics sets in and both observables show a clear power-law 
behavior. We found an exponent for the imbalance which is the double of the one for the correlation function, consistently with 
what found for many-body systems and expected on the basis of scaling arguments~\cite{BarLev,reviewdeloc1}. By increasing 
$W$ and approaching the localization transition the exponents vanish. These results, which are strikingly similar to the ones found approaching the MBL transition in simulations and experiments~\cite{dave1,demler,alet,torres,BarLev,reviewdeloc1,experiments1,experiments2,experiments3}, emerge in the part of the phase diagram where previous analysis of the spectral properties have suggested the presence of a non-ergodic delocalised 
phase~\cite{altshuler}. \\
In order to establish a direct link between the two phenomena it is useful to introduce the spectral probe~\cite{chalker,kravtsov2,BarLev2}: 
\[
K_2(E)= \overline{\frac{1}{\mathcal{N}}\sum_{\alpha,\beta}'\sum_{x=1}^M|\langle \alpha|x\rangle|^2
|\langle \beta|x\rangle|^2 \delta[E-(E_\beta-E_\alpha)]} \, ,
\]
where $\mathcal N$ is a normalization factor (chosen in such a way that the integral over $E$ is equal to one). 
This function was introduced to characterize the critical properties of the Anderson localization transition  
and recently studied to probe the putative non-ergodic delocalized phase~\cite{kravtsov,altshuler,kravtsov3}. We show its behavior in Fig.~\ref{fig:pd} for different system sizes and $W=14$. In a standard ergodic delocalized phase $K_2(E)$
converges in the thermodynamic limit to a function with a finite value at zero energy \cite{kravtsov,kravtsov2}. This is indeed what happens for very small values of $W$; for $W=0$ we fully recover the GOE behavior. 
In the interval $10 \lesssim W \lesssim W_c$ the function $K_2(E)$ instead displays an apparent power-law regime that extends to smaller and smaller energy the larger is the systems size. Since $K_2(E)$ can be 
related to our proxy for the equilibrium correlation function through $C(t) \sim \int {\rm d} E K_2(E) \cos(E t)$, the power-laws observed in Figs.~\ref{fig:Ct} and \ref{fig:IB} are 
actually tightly linked to the power-laws in energy found for $K_2(E)$. A direct numerical proof is that the power-law exponent 
of the imbalance, $\beta$, of the equilibrium correlation function, $\alpha$, and of $K_2(E)$, $\mu$, are all related within numerical accuracy: $\alpha\simeq \beta/2\simeq 1-\mu$, see inset of Fig.~\ref{fig:pd}. 
All this clearly shows that the physical origin of the unusual time dependence observed in the correlation functions is the non-ergodic-like features of the spectral statistics that emerge within the delocalized phase before that the transition takes place.\\
As already stressed, the existence of a {\it bona-fide} non-ergodic delocalized phase has recently been matter of debate. All the crux of showing the presence of such intermediate phase is finding out whether or not in the $M \to \infty$ limit the power-law divergence in $E$ is cut-off for a certain (very small) value $E_c$ or instead it persists forever. In the latter case 
the power-laws in time would extend for arbitrary large times, whereas in the former they would be cut-off 
at very long times. Whatever is the correct answer, our results show that on a very large time-window the system behaves for all practical purposes {\it as if} there were a non-ergodic delocalized phase, and that this explains the power-law behavior of physical correlation functions. \\
The existence of a cross-over size, $M_c$, above which one would find back normal ergodic behavior has received some evidences from analytical and numerical works~\cite{mirlin,lemarie,levy}; approaching the transition the logarithm of $M_c$ is expected to diverge as $1/\sqrt{W-W_c}$~\cite{susy}. 
It is natural to assume that the characteristic energy $E_c$ scales as the inverse of the correlation volume $M_c$ dominating the finite size effects, as it happens for ordinary Anderson localization transitions~\cite{kravtsov,kravtsov2}, thus implying that the power-law dependence and slow dynamics are cut-off on a time-scale $\tau(W)\sim e^{c/\sqrt{W-W_c}}$, where $c$ is a positive constant. 
If this were the case then the dynamics of the system would be very slow and unusual 
on a very large time-window spanning many decades, although it might become eventually ergodic---a situation that strikingly resembles the one of super-cooled liquids close to the glass transition. Taking inspiration from this case, as also suggested in~\cite{altshuler}, we dubb glassy and delocalized the regime before $W_c$. Establishing whether this regime is truly non-ergodic is a very interesting theoretical question but irrelevant for many practical purposes (as also is the existence of a {\it bona-fide} ideal glass transition). Moreover, note that what leads to the power-laws we found in the imbalance and the correlation functions is 
the behavior of $K_2(E)$ at arbitrary small but finite energies. This relates unusual slow dynamics to 
unusual spectral properties on the scale $E_c$, and not on the scale of the mean level-spacing $\delta$ ($E_c\gg \delta$ even in presence of an intermediate phase \cite{kravtsov}), in agreement with recent results 
\cite{BarLev2,kravtsov3,torres}. \\
Let us finally comment on the relationship with Griffiths effects in real space. Inclusions of the localised phase in real space lead to kinetic bottlenecks in Fock space and, consequently, to a very heterogenous many-body delocalisation~\cite{deroeck}. Similarly, on the Bethe lattice, the unusual properties of the delocalised phase have been traced back to delocalisation along rare paths. In this case the difference between the ergodic and non-ergodic phases is very similar to the one between the normal and glassy phases of directed polymers in random media, where preferred conformations of the polymer correspond to favorable delocalization paths~\cite{noi,altshuler,mirlin,lemarie} 
(these non-ergodic features might extend only up to a large but finite length-scales if the delocalized phase indeed becomes standard and fully ergodic asymptotically). 
The interest of focusing on real space to explain the unusual dynamics of the delocalized phase is that this leads to a very intuitive and concrete theoretical explanation.  
The advantage of focusing on Fock space is instead that this perspective provides a more general framework that could be relevant also in cases where, strictly speaking, Griffiths effects do not hold or should be milder~\cite{griffiths2}, such as in $2d$ disordered or $1d$ quasi-periodic systems, but the dynamics is slow and unusual as for $1d$ disordered systems~\cite{experiments3,daverecent,dave2d}. Understanding the precise connection between the real space and the Fock space perspectives certainly requires to go beyond the simple toy model studied in this work. This is something worth future investigations for which our results provide useful guidelines.

\begin{acknowledgments}
We thank Y. Bar-Lev, D. Reichman, V. Ros, M. Schiro, A. Scardicchio for helpful discussions and
acknowledge support from the ERC grant NPRGGLASS and by a grant from the Simons Foundation (\#454935, Giulio Biroli).
\end{acknowledgments}


\newpage 
\newpage
\section{Supplemental Material}

In this supplemental materials we illustrate and discuss in more detail the finite-size effects observed on the time-dependence of dynamical observables at small sizes 
and very large times-scales.\\

\begin{figure}[h]
\includegraphics[width=0.42\textwidth]{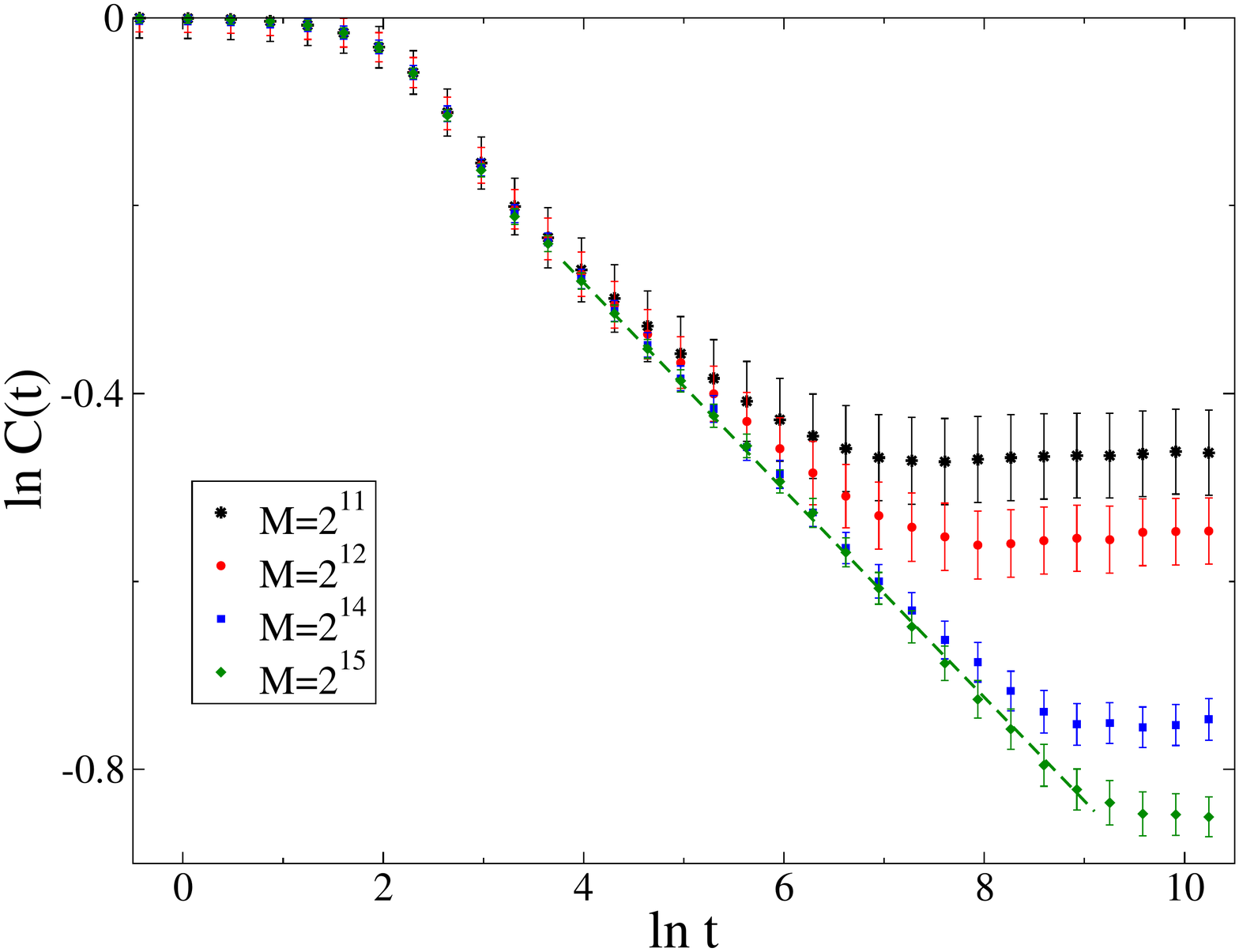}%
\hspace{1cm}
\includegraphics[width=0.42\textwidth]{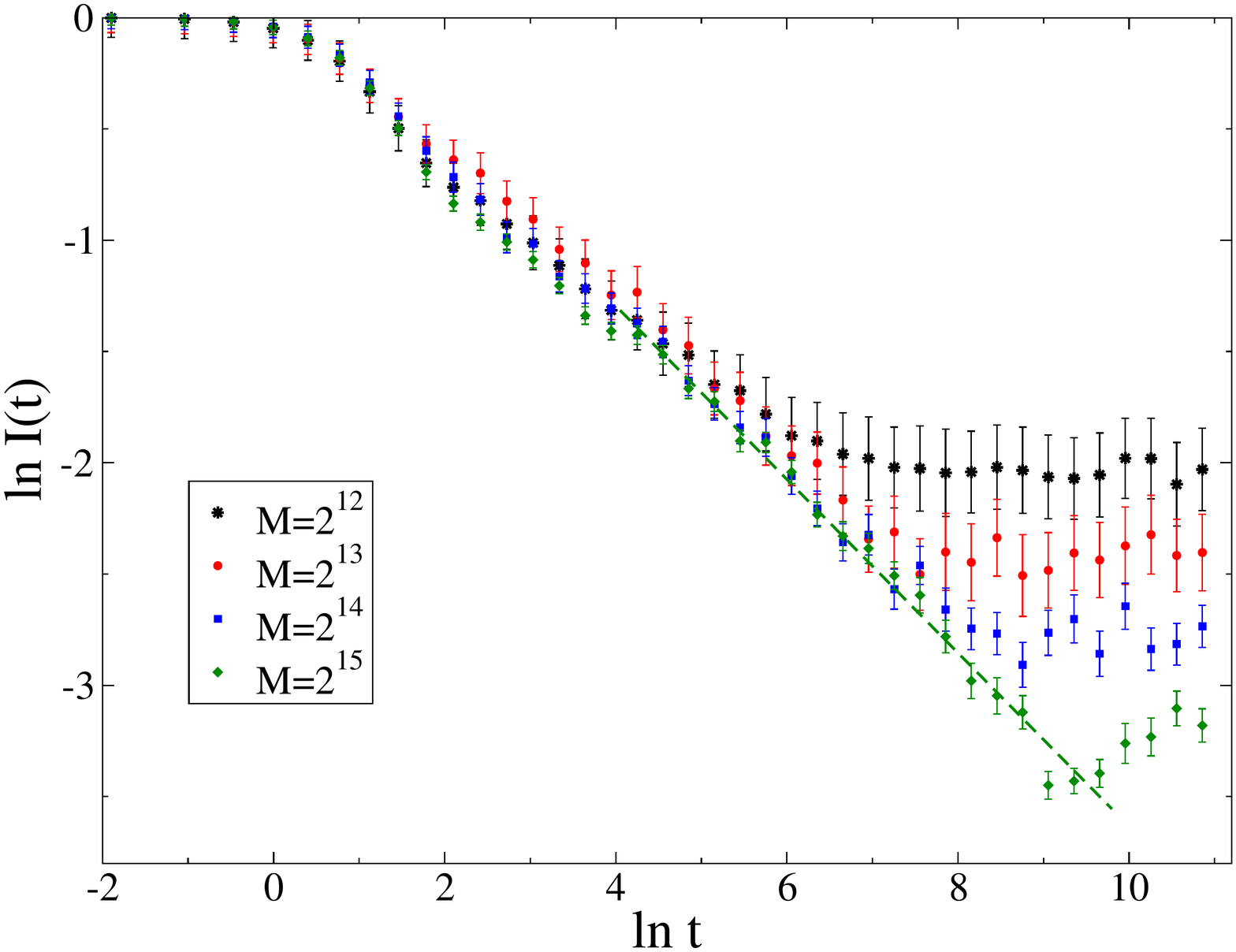}
\caption{\label{fig:W14}
Equilibrium correlation function $C(t)$ for $W=14$ (top) and imbalance $I(t)$ for $W=13$ 
(bottom) 
as a function of time and different system sizes $M=2^n$ with $n=11, \ldots 15$ (log-log plot). 
The dashed straight lines highlight the apparent power-law behaviors, $C(t)\sim t^{-\alpha}$ and $I(t)\sim t^{-\beta}$.} 
\end{figure}

In Fig.~\ref{fig:W14} we plot the time dependence of the equilibrium correlation function $C(t)$ for $W=14$, of the imbalance $I(t)$ for 
$W=13$, and for different system sizes $M=2^n$ with $n=11, \ldots, 15$. The data are averaged over $2^{21-n}$ realizations of the disorder.
After a relatively fast decay at short times, a slow power-law regime sets in for intermediate times, where the dynamical observables behave
as $C(t) \sim t^{-\alpha}$ and $I(t) \sim t^{-\beta}$. 
At large $t$ the curves depart from the straight lines and approach a plateau value whose height decreases as the system size is increased. 
As a result, the time window over which the power-law decays are observed gets broader for larger samples.
The exponents $\alpha$ and $\beta$ of the power-laws appear however to be independent on the system size.\\
The key observation is that the finiteness of $M$ only affects the long-time value on scales that are larger the larger is $M$.   
The plateau values and their scaling with $M$ are related to the possible multi-fractal behavior.  
For instance, the plateau value of the correlation function equals the Inverse Participation Ratio,
defined as $\Upsilon_2 \equiv \overline{\sum_x | \langle \alpha | x \rangle |^4}$ 
(where the average is performed over the eigenstates around zero energy). Its scaling with $M$ 
is $\Upsilon_2 
\sim M^{-D_2}$ where $D_2$ is a fractal dimension. Likewise, one finds that the plateau 
of the imbalance goes to zero with $M$ as a power law. 
If the system is ergodic the exponents of these power-laws are trivial, e.g. $D_2=1$. \\
In conclusion, only the very large time behaviour (diverging with $M$) suffers of finite size effects. The plot of the imbalance and the correlation function presented in the main text, which are for large but finite times, can be considered converged for $M\rightarrow \infty$. Put it in more mathematical terms, pointwise convergence  is reached for not too large values of $M$ even for relatively large times. \\

Our numerical findings show that at  smaller values of $W$, for instance, at $W=10$, the incipient power-law decay observed at intermediate times is cut-off and replaced at very large times by a 
faster (probably stretched exponential) decay, which can be observed only for the largest accessible system sizes ($M = 2^{15}$).
For $11 \lessapprox W < W_c$ the bigger the system size the longer the power-law regime holds. In order to 
find out whether there is cut-off in this regime, one should reach very long time-scales and, hence, beat finite size effects, that induce a spurious plateau, by using very large 
system sizes $M$. 


\begin{thebibliography}{99}
\bibitem{Bethe_dinf} In very large dimensional lattices the number of neighbors increases very fast with distance and loops are suppressed. Both features are well encoded in Bethe lattices which display a local tree-like structure.
\bibitem{diagonal} 
For the sake of simplicity we restrict the analysis  to observables diagonal in the $|{\cal C}\rangle$ basis. Extending it to non-diagonal ones is straightforward and not expected to bring any additional insight.
\bibitem{microcanonical_Tinf} Restricting around zero energy means that we focus on a small interval $[-\epsilon,+\epsilon]$ where 
$\epsilon$ is small on scales of the bandwidth ($\sim W$) but it is much larger than the scale of the mean-level spacing which is of the order $1/M$. In practice we have taken $M/16$ energy levels around $E=0$ which corresponds to $\epsilon \approx 0.47$ for $W=14$. We have verified that 
the our results do not change if the interval is reduced by a factor $4$ or increased by a factor $2$.
\bibitem{average} The averages are performed over the disorder (\i.e., different realizations of the
on-site random energies and of the random-regular graph) and the random coefficients of $f_{\sigma_i^z} (x)$.
For the imbalance $I(t)$ we also average over the initial conditions $|x_0\rangle$.
\end{thebibliography}
\end{document}